\documentclass[epj]{svjour}
\usepackage{psfig}
\begin{document}
\title{Hadrons with charmed quarks in matter}
\author{A. Sibirtsev}
%
\institute{Institut f\"ur Kernphysik, Forschungszentrum J\"ulich,
D-52425 J\"ulich}
\date{Received: date / Revised version: date}
%
\abstract{
We investigate the $D{\bar D}$ decay width of 
excited charmonium states at finite nuclear density with 
simultaneous modification of both $D$ and ${\bar D}$ mesons in 
nuclear matter. The strongest effect is found for the $\Psi^{\prime}$ meson.
The medium modification can be detected by dilepton spectroscopy 
as substantial $\Psi^{\prime}$ broadening and anomalous $\Psi^{\prime}$
absorption.
} 
\maketitle
\vspace*{-5mm}\section{Introduction.}
In dense and hot nuclear matter the light quark condensates $q{\bar q}$
might be substantially reduced. This affects the light quark content 
of mesons and baryons and therefore results in medium modification  of 
hadron properties~\cite{Hatsuda,Weise,Tsushima1}.
Even if the changes in quark condensates are small, the absolute
difference between the in-medium and bare  masses of  hadrons is
expected~\cite{Tsushima1,Sibirtsev1} to be larger for heavier hadrons.

\begin{figure}[b]
\vspace*{-6mm}
\hspace*{-1mm}\psfig{file=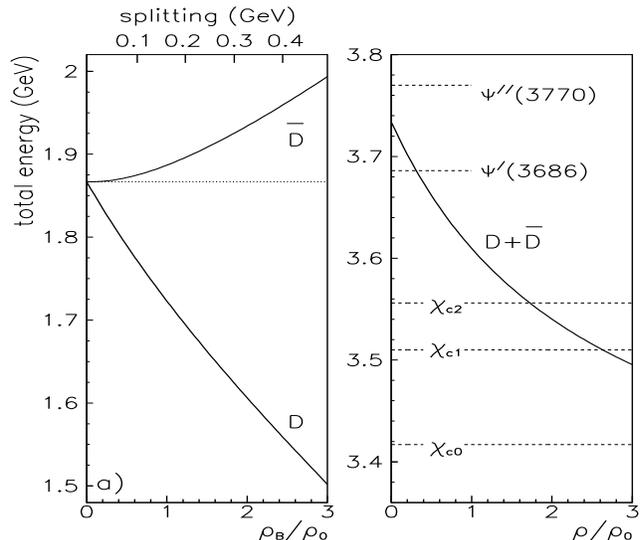,width=8.9cm,height=7.6cm}
\vspace*{-8mm}
\caption{a) The in-medium mass of $D$ and ${\bar D}$ mesons as a function
of nuclear density $\rho{/}\rho_0$, with $\rho_0$=0.16~fm$^{-3}$
and the $D{\bar D}$ mass splitting (upper axis). b) The solid line
shows overall $D{\bar D}$ mass as a function of density.
The dashed lines indicate the masses of excited charmonia.}
\label{comic7}
\end{figure}

The charmed mesons, which consist of light $q,{\bar q}$
and heavy $c,{\bar c}$ quarks, are
considered suitable probes of in-medium modification of
hadron properties. Similaly to ${\bar K}$ (${\bar q}s$) and
$K$ ($q{\bar s}$) mesons, the $D$ (${\bar q}c$) and $\bar{D}$ ($q{\bar c}$)
satisfy different dispersion relations in matter because
of the different sign of $q$ and ${\bar q}$ vector 
couplings~\cite{Tsushima1}. While the $D$ meson mass is reduced 
in nuclear matter, the ${\bar D}$ mass is raised, as is illustrated by 
Fig.\ref{comic7}a). Calculations with the Quark Meson Coupling 
model (QMC)~\cite{Guichon} show that already at normal nuclear density 
$\rho_0$ the mass splitting between  $D$ and ${\bar D}$ mesons 
is about 160~MeV. 

It was proposed~\cite{Sibirtsev1} that the modification of the $D$ 
meson in nuclear matter can be identified by enhanced subthreshold 
production of open charm in ${\bar p}A$ annihilation. Because
of charm conservation, $D$ and ${\bar D}$ mesons are produced 
pairwise. As is shown  in Fig.\ref{comic7}b) the sum of 
$D$ and ${\bar D}$ masses  depends substantially
on nuclear density. The downward shift of the $D{\bar D}$ 
threshold at $\rho_0$ is $\simeq$--100~MeV. A QCD 
sum rule estimate~\cite{Hayashigaki} predicts about the 
same shift.

Furthermore, an attractive $D$-nucleus potential can be measured by 
investigating charmed mesic nuclei~\cite{Tsushima1}. The 
reduction of $D$ mass in matter might affect open charm 
production~\cite{Cassing} and $J{/}\Psi$ suppression~\cite{Sibirtsev2} 
in relativistic heavy ion collisions.

In contrast to open charm, charmonium mesons consist of
heavy $c{\bar c}$ quarks and can be affected only by gluonic 
condensates. It was expected that the modification of properties 
of heavy quarkonia in matter would be almost negligible~\cite{Brodsky}.
In that case, the overall $D{\bar D}$ mass 
at some nuclear density might cross the masses of excited 
charmonium states, as is illustrated by Fig.\ref{comic7}b). 

The crossing of the charmonium states levels and the $D{\bar D}$
threshold in nuclear medium might result in the
melting~\cite{Digal} of excited charmonium mesons.

Here we investigate the modification of widths of charmonium
states at finite nuclear density with $D$ and ${\bar D}$
in-medium masses predicted by QMC~\cite{Tsushima1}.  We consider 
simultaneous modification of both $D$ and ${\bar D}$ mesons, 
as is illustrated in Fig.\ref{comic7} and study 
$\Psi^{\prime\prime}$(3770), 
$\Psi^{\prime}$(3686) and $\chi_{c2}$ decay into
$D{\bar D}$ in nuclear matter. The $\chi_{c0}$
modification is not discussed since the $D{\bar D}$ threshold
does not cross its mass even at $\rho$=3$\rho_0$, as is shown in
Fig.\ref{comic7}. The $\chi_{c1}{\to}D{\bar D}$ decay is 
suppressed by parity conservation.

\vspace*{-2mm}\section{$\Psi^{\prime\prime}$(3770).}
The $\Psi^{\prime\prime}$ charmonium lies above the $D{\bar D}$ threshold
in free space and its dominant decay width into $D{\bar D}$ channel is 
given by
\begin{equation}
\Gamma_{\Psi^{\prime\prime}{\to}D{\bar D}}{=}
\frac{g_{\Psi^{\prime\prime}{D{\bar D}}}^2}{3\pi}
\frac{q^3}{m_\Psi^2},
\label{eq1}
\end{equation}
where $m_\Psi$ is the $\Psi^{\prime\prime}$(3770) mass and $q$ is 
the $D$ meson momentum in the charmonium rest frame,
\begin{equation}
q=\frac{[(m_\psi^2{-}m_D^2{-}m_{\bar D}^2)^2{-}4m_D^2m_{\bar D}^2]^{1/2}}
{2m_\Psi},
\end{equation}
with $m_D$ and $m_{\bar D}$ the  masses of
$D$ and ${\bar D}$ mesons, respectively, while the coupling  constant
$g_{\Psi^{\prime\prime}{D{\bar D}}}$=14.89   is fixed by the 
vacuum  decay width $\Gamma_{\Psi^{\prime\prime}{\to}D{\bar D}}$=23.6 MeV. 

\begin{figure}[b]
\vspace*{-13mm}
\hspace*{-2mm}\psfig{file=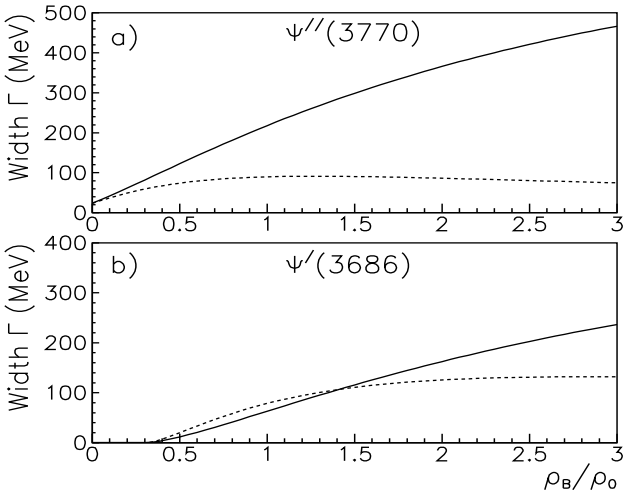,width=9.cm,height=8.cm}
\vspace*{-9mm}
\caption{Decay widths of $\Psi^{\prime\prime}(3770)$ a) and 
$\Psi^\prime(3686)$ b) charmonium 
states into $D{\bar D}$ as a function of nuclear matter
density in units of $\rho_0$ . The solid lines show the phase space 
dependence, while the
dashed lines are the results from the $^3P_0$ model. In both calculations 
the $D$ and ${\bar D}$ in-medium masses are given by the QMC model.}
\label{comic7f}
\end{figure}

If $\Psi^{\prime\prime}{D{\bar D}}$ coupling does not change in matter 
the modification of the $\Psi$(3770)  width is entirely given by
$D$ and ${\bar D}$ in-medium masses and is determined by the
phase space dependence of the charmonium decay width. In that case the 
in-medium $\Psi$(3770) width  depends substantially on nuclear matter 
density, as is shown by the solid line in Fig.\ref{comic7f}a).

Within the $^3P_0$ model~\cite{Busetto} the $\Psi^{\prime\prime}{D{\bar D}}$ 
coupling itself depends on the $D$ and ${\bar D}$ masses via
\begin{eqnarray}
g^2_{\Psi^{\prime\prime}{D{\bar D}}}{=}\frac{\pi^{3/2}\, 2^{11}\, 
\gamma^2m_\Psi}
{ 5\, \beta^7 \, 3^{10}}\,\, [(q^2+m_D^2)(q^2+m_{\bar D}^2)]^{1/2}
\nonumber \\
\times(15\beta^2-2q^2)^2 \, \exp{(-q^2/6\beta^2)},
\end{eqnarray}
where the oscillator length scale $\beta$=360~MeV is fixed by 
light mesons decays~\cite{Busetto}, while the interaction strength
$\gamma$=0.33 is determined by $\Psi^{\prime\prime}{\to}D{\bar D}$
decay.

Finally, the dashed line in Fig.\ref{comic7f} shows the dependence of 
the $\Psi^{\prime\prime}{\to}D{\bar D}$ decay width on nuclear matter 
density resulting from the $^3P_0$ model. At normal nuclear density 
$\rho_0$= 0.16 fm$^{-3}$ the $\Psi^{\prime\prime}(3770)$ in-medium 
width almost saturates  
$\Gamma_{\Psi^{\prime\prime}{\to}D{\bar D}}{\simeq}$90~MeV,
which is $\simeq$3.8 times larger than in vacuum. Furthermore, the
$^3P_0$ result substantially differs from the phase space
estimate.

The $\Psi^{\prime\prime}$ modification in nuclear matter might be
studied by dilepton spectroscopy from $AA$
as well as ${\bar p}A$ interactions, since
the effect is measurable already at normal nuclear densities.
The $\Psi^{\prime\prime}(3770)$ charmonium does not decay into $J{/}\Psi$
and thus its modification in matter can not be considered as 
an additional source of $J{/}\Psi$ suppression in heavy ion 
collisions.

\vspace*{-2mm}\section{$\Psi^\prime$(3686).}
The $\Psi^\prime$ charmonium lies below the $D{\bar D}$ threshold
in free space, but the $D{\bar D}$ decay channel becomes open 
at nuclear density $\rho{\simeq}$0.05~fm$^{-3}$, as  illustrated
by Fig.\ref{comic7}. The $\Psi^\prime$ is narrow, its total width
is 0.277~MeV and partial decay into $J{/}\Psi$  accounts for 
$\simeq$54\% of the total width. 

Although, the $\Psi$(3686) coupling to $D{\bar D}$ is not directly
accessible, it can be estimated within the framework of
Vector Meson Dominance model as
\begin{equation}
g^2_{\Psi^\prime{D{\bar D}}}{=}\frac{16 \pi \alpha^2}{27}\frac{m_\Psi}
{\Gamma_{\Psi^\prime{\to}e^+e^-}},
\end{equation}
where $\alpha$ is the fine structure constant, $m_\Psi$ is the
$\Psi^\prime$(3686) mass and $\Gamma_{\Psi^\prime{\to}e^+e^-}$=2.35~KeV 
is the radiative $\Psi^\prime{\to}e^+e^-$ decay width. Finally, 
the $\Psi^{\prime}D{\bar D}$ coupling constant is 
12.84;  the $\Psi^{\prime\prime}D{\bar D}$
coupling from VMD is to 19.94, which is close to the 
result from direct $\Psi^{\prime\prime}{\to}D{\bar D}$ decay 
given by Eq.\ref{eq1}.

The phase space dependence of the $\Psi^\prime$ in-medium  width from
Eq.\ref{eq1} is shown by the solid line in Fig.\ref{comic7f}b).
This result again can be compared with the prediction of the $^3P_0$
model given by~\cite{Busetto}
\begin{eqnarray}
\Gamma_{\Psi^\prime{\to}D{\bar D}}
{=}\frac{\pi^{1/2}\, 2^9\, \gamma^2}
{m_\Psi\, \beta^7 \, 3^{11}}\,\, [(q^2+m_D^2)(q^2+m_{\bar D}^2)]^{1/2}
\nonumber \\
\times q^3\, (15\beta^2-2q^2)^2 \, \exp{(-q^2/6\beta^2)}
\end{eqnarray}
and shown by the dashed line in Fig.\ref{comic7f}b). Here the calculations
were done with parameters $\beta$ and $\gamma$ evaluated above.
Note that at nuclear densities
$\rho{\le}1.5\rho_0$ both the $^3P_0$ model and VMD phase space estimates
are in reasonable agreement. An exciting observation is that
at normal nuclear density $\rho_0$=0.16 fm$^{-3}$ the $\Psi^\prime$ 
width is $\simeq$70~MeV, which is $\simeq$250 times larger 
than that in vacuum.

Again, the $\Psi^\prime$ modification in nuclear matter can be
measured through dilepton spectroscopy. As we found, the 
dilepton spectrum from $\Psi^\prime(3686)$ decay in matter
might strongly overlap with $\Psi^{\prime\prime}(3770)$ 
charmonium decay. Fig.\ref{comic7g}a) shows dilepton spectra from
$\Psi^\prime$ and $\Psi^{\prime\prime}$ decays unfolded from the
production cross section. The solid lines show the results for
matter at normal nuclear density, while 
the dashed lines indicate the spectra in free space. 
The measurement of a broad peak around $\Psi^\prime$ pole might be
considered as an direct indication of in-medium modification
of charmonia.

On the other hand, a large $D{\bar D}$ component of $\Psi^\prime$ charmonium
should result in strong $\Psi^\prime$ absorption in nuclear matter,
similar to that found~\cite{Sibirtsev2} for $J{/}\Psi$ dissociation.
Moreover, $\Psi^\prime$ melting in nuclear matter will additionally suppress 
$\Psi^\prime{\to}J{/}\Psi$ decay and partially eliminate $J{/}\Psi$
yield in heavy ion collisions.

\vspace*{-2mm}\section{$\chi_{c2}$(3556).}
The $\chi_{c2}$  charmonium level crosses the $D{\bar D}$
threshold at a nuclear density of about 0.28~fm$^{-3}$. 
In vacuum the $\chi_{c2}$
width is 2~MeV and  partial decay into the $J{/}\Psi$  is
$\simeq$13.5\%. There is no reliable way to estimate $\chi_{c2}D{\bar D}$ 
coupling and to evaluate the phase space dependence of 
the $\chi_{c2}$ in-medium width. 

\begin{figure}[b]
\vspace*{-14mm}
\hspace*{-3mm}\psfig{file=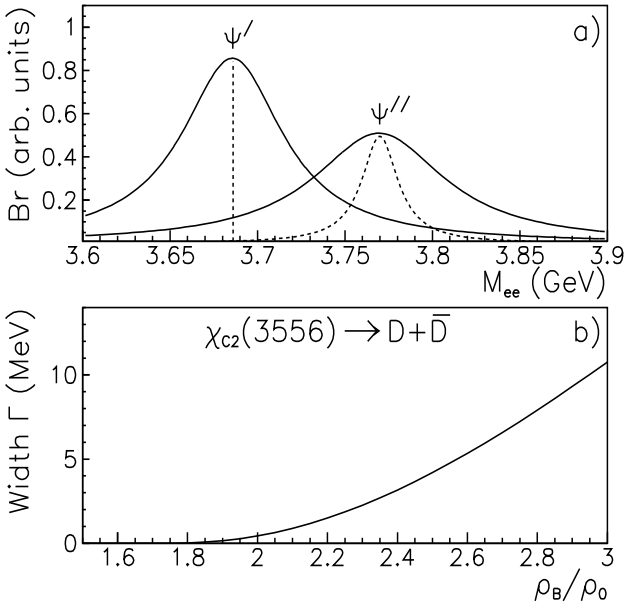,width=9.cm,height=9.8cm}
\vspace*{-6mm}
\caption{a) Dilepton spectra from $\Psi^\prime$ and $\Psi^{\prime\prime}$
decays at normal nuclear density $\rho{=}\rho_0$=0.16~fm$^{-3}$
(solid lines) and in vacuum (dashed lines). b) Decay width of 
$\chi_{c2}(3556)$  charmonium  into $D{\bar D}$ as a 
function of nuclear matter density in  units of  $\rho_0$ predicted by
the $^3P_0$ model with  $D$ and ${\bar D}$ in-medium masses given by QMC.}
\label{comic7g}
\end{figure}

In the $^3P_0$ model the $\chi_{c2}{\to}D{\bar D}$
width is given as~\cite{Busetto}
\begin{eqnarray}
\Gamma_{\chi_{c2}{\to}D{\bar D}}
{=}\frac{\pi^{1/2}\, 2^{12}\, \gamma^2}
{5 \, m_\chi \, \beta^5 \, 3^{8}}\,\, [(q^2+m_D^2)(q^2+m_{\bar D}^2)]^{1/2}
\nonumber \\
\times q^5 \, \exp{(-q^2/6\beta^2)},
\end{eqnarray}
where $m_\chi$ is the $\chi_{c2}$ mass and parameters $\beta$ and 
$\gamma$ are listed above. The in-medium $\chi_{c2}$ width is shown in
Fig.\ref{comic7g}b) as a function of nuclear matter density.
The $\chi_{c2}$ modification becomes significant only at large 
densities.

\vspace*{-2mm}\section{Conclusion.}
Modification of the $D$ and ${\bar D}$ meson masses in nuclear 
matter leads to a substantial increase of the $\chi_{c2}$,
$\Psi^\prime$ and $\Psi^{\prime\prime}$ decay widths into
the $D{\bar D}$ channel. 

The calculations with the density independent coupling constants 
between the $\Psi^\prime$ and $\Psi^{\prime\prime}$ 
charmonium and $D{\bar D}$ 
pair results in strong and monotonic density dependence of 
the $\Psi^\prime$ and $\Psi^{\prime\prime}$ in-medium widths due 
to increase of the final state phase space. Within the
$^3P_0$ model these couplings are also considered as a function 
of the in-medium  $D$ and ${\bar D}$ masses and as a result the
$\Psi^\prime$ and $\Psi^{\prime\prime}$ charmonium widths
do not increase monotonically with nuclear density, but
saturate at $\rho_0$=0.16~fm$^{-3}$. It was found that
the saturation limits are 
$\Gamma_{\Psi^\prime{\to}D{\bar D}}{\simeq}$70~MeV and
$\Gamma_{\Psi^{\prime\prime}{\to}D{\bar D}}{\simeq}$90~MeV,
which can be compared with vacuum widths of 0.277~MeV 
and 23.6~MeV, respectively. 

The $\chi_{c2}$  charmonium level crosses the $D{\bar D}$
threshold at $\rho{\simeq}$ 1.25$\rho_0$ and its
width increases with nuclear matter density.
At $\rho{\simeq}$3$\rho_0$ the $\chi_{c2}$ decay width 
into $D{\bar D}$  is about 11~MeV, which may be compared with
total $\Gamma_{\chi_{c2}}$=2~MeV width in vacuum.

We conclude that  modification of the $D$ and ${\bar D}$ in matter  
most dramatically affects the $\Psi^\prime$ charmonium.
This can be detected by dilepton spectroscopy as
the appearance of a broad peak near the $\Psi^\prime$ pole or as
anomalous $\Psi^\prime$ suppression in nuclear  matter.

Our results are in agreement 
with the most recent findings~\cite{Ko} 
that both the mass and width of $\Psi^{\prime\prime}$(3770) 
depend significantly on the $D$ meson mass.

\end{document}